\documentclass[nofootinbib,amsmath,amssymb,pdftex]{revtex4}
\usepackage{amsmath}
\usepackage{graphicx}
\usepackage{dcolumn}
\usepackage{bm}
\def\l{\lambda}
\def\t{\theta}
\def\z{\zeta}
\def\hs{\hspace}
\begin{document}

\title{Transition Radiation from the Neutrino-Photon Interaction in Matter}

\author{Juan Carlos D'Olivo}
\email{dolivo@nucleares.unam.mx}
\author{Jos\'e Antonio Loza}
\email{aloza@ibt.unam.mx}
\affiliation{Instituto de Ciencias Nucleares,Universidad Nacional
Aut\'{o}noma de M\'{e}xico,\\ Apartado Postal 70-543, 04510
M\'{e}xico, D.F., Mexico}

\begin{abstract}
We show that, because of their effective electromagnetic interaction in matter,
transition radiation is emitted whenever neutrinos goes across the boundary between 
two media with different indices of refraction. This effect occurs in the context of the
standard model and does not depend on any exotic neutrino property.
We examine such a phenomena and  compare it with the transition radiation of a neutrino 
endowed with an intrinsic dipole moment.  
\end{abstract}

\maketitle

The electromagnetic properties of neutrinos are of
great relevance in a variety of physical, astrophysical, and
cosmological contexts \cite{giunti}. From the experimental side,
up to now there is no evidence confirming a nonzero value for any 
of these properties \cite{raffelt}. Any observed deviation from the predictions 
of the standard model (SM), minimally extended to accommodate neutrino masses, 
would have a profound implication for the search of new physics \cite{bell, novales}.
The electromagnetic characteristic of the neutrinos can also serve to elucidate
whether they are Dirac or Majorana fermions \cite{diracmaj}. 
On the other hand, as is now well known, the basic properties of neutrinos  
that propagate through a medium can be substantially
different compared to their properties in the vacuum. 
In particular, because of their weak interactions with the charged leptons and 
nucleons in a background, neutrinos acquire an effective coupling to the electromagnetic field
\cite{sem, dn}. This fact can give rise to several interesting physical
processes: plasmon decay \cite{plasmon,zaidi}, absorption 
of electromagnetic waves \cite{tsytov}, radiative neutrino decay \cite{radiative}, 
and Cherenkov radiation by chiral neutrinos \cite{sawyer, cherneu}. 
\begin{figure}[h]
\label{figGeom}
\includegraphics[viewport=20 24 288 130, width= 11 cm]{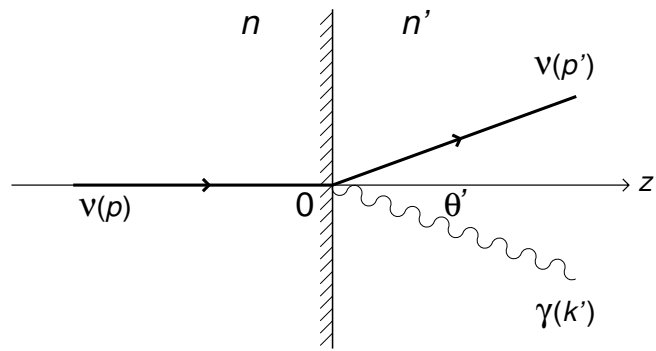}
\caption{Transition radiation at the interface between a medium with refraction
index $n$ and the vacuum ($n' = 1$).}
\end{figure}

In a uniform medium, the radiative process
\begin{equation}
\label{process} \nu(p)\rightarrow \nu(p')+ \gamma(k)
\end{equation}
is kinematically allowed if $n > 1$ and $v > 1/n$, where $n$ is the
refraction index of the medium and $v$ the neutrino velocity. This
leads to the Cherenkov radiation. There exists another important
radiative process that can take place even if the above conditions
are not satisfied: the transition radiation (TR) \cite{frankg}. Such
radiation is emitted whenever a charged particle goes across the
boundary between two media with different indices of refraction.
The phenomenon also happens with a neutral particle having  
a non vanishing dipole moment. The TR of a neutrino produced by  
an intrinsic (magnetic, electric, or toroidal) dipole moment
have been examined by several authors \cite{sakuda, grimus, bukina} and 
an application of such effect as a new technique to measure the 
neutrino magnetic moment has been proposed by Sakuda \cite{sakuda}.

As was pointed out several years ago \cite{cherneu}, when neutrinos cross 
the interface of two media they emit TR because of their effective electromagnetic 
interaction in matter \cite{doloza}. To our knowledge this process has not received 
the attention it deserves,  despite the fact that is a novel physical prediction that
does not hinge on hypothetical neutrino properties and/or interactions beyond
those of the SM. Hence, it has to be there and needs to be properly characterized 
in order to distinguish it from a similar effect that could be associated with 
new physics. In this paper we present a quantum theoretical calculation of the 
TR due to the neutrino-photon coupling in an electron background. As we show,
the effect under consideration can be comparable, and even larger, than the one 
due to an intrinsic electromagnetic characteristic of the neutrino, and therefore
should be taken into account  when analyzing a TR experiment eventually 
designed to measure the neutrino magnetic moment.

Let us consider a neutrino beam crossing the plane interface between
a material medium and the vacuum, as shown in Fig. 1. For
definiteness, we restrict ourselves to the case where neutrinos go
from the medium into the vacuum, but a similar calculation can be
done for the reverse situation or the case where instead of the
vacuum we have another material medium. We choose the coordinate
system in such a way that, in the frame where the medium is at rest,
the interface coincides with the plane at $z=0$ and the medium
occupy the region located at $z<0$. In this frame, $p^{\,\mu}=
({\mathcal E},\boldsymbol{\wp})$ and $p^{\,\prime\mu} =
({\mathcal E}^{\prime},\boldsymbol{\wp'})$ are the components of the
initial and final momenta of the neutrino. The incident neutrinos
are assumed to move along the $z$ axis, perpendicularly to the
interface, that is $\boldsymbol{\wp} = \wp\,\mathbf{\hat{z}}$, with 
$\wp= |\boldsymbol{\wp}|$.\,The four momentum of the emitted photon is 
$k^\mu =(\omega, \boldsymbol{\kappa})$ in the medium and
$k'^\mu=(\omega', \boldsymbol{\kappa'})$ in the vacuum, with $\omega =
\omega'$ but $\boldsymbol{\kappa}\neq\boldsymbol{\kappa'}$ due to 
the nonconservation of the momentum along the $z$ direction.

The influence of the medium on the energy momentum relation of the
neutrinos does not play an essential role in our analysis and we take
$p^{\,2} = p^{\,\prime\,2} = m^2_\nu $, where $m_\nu$ is the
neutrino mass.  On the contrary, in general it is not correct to disregard
the matter effects on the photon dispersion relation
$\omega(\kappa)$. They can be determined from the background
contributions to the polarization tensor $\pi_{\mu\nu}$. In an
isotropic medium \cite{weldon,nipal}
\begin{equation}
\label{fotonselfe} \pi_{\mu\nu}(k) = \pi_{\!_T} R_{\mu\nu} +
\pi_{\!_L} Q_{\mu\nu} \,,
\end{equation}
where
\begin{eqnarray}
\label{R}
R_{\mu\nu} &=& g_{\mu\nu} - \frac{k_\mu k_\nu }{k^2} - Q_{\mu\nu} \,, \\
\label{Q} Q_{\mu\nu} &=& - \frac{k^2}{\kappa^2} \left( u_\mu -
\frac{\omega}{k^2}\,k_\mu\right) \left( u_\nu - \frac{\omega}
{k^2}\,k_\nu \right) ,
\end{eqnarray}
are mutually orthogonal tensors ($R_{\mu\nu}Q^{\mu\nu}=0$) that
satisfy the relations $R_{\mu\nu}R^{\mu\nu}=2$ and $ Q_{\mu\nu}
Q^{\mu\nu}=1$. Here,
$u^\mu=(1, \bf{0})$ is the four velocity of the medium, while the
coefficients $ \pi_{\!_{T,L}}(\omega,\kappa)$ are scalar functions
of the invariant quantities $\omega=k \cdot u$ and 
$\kappa=\sqrt{\omega^2 - k^2}$, with $\kappa = \lvert\boldsymbol{\kappa}\rvert$. 

Since the longitudinal mode (plasmon) does not propagate in vacuum, only 
the two degenerated transverse modes (photon) contribute to the processes we 
are interested in. The photon frequency within the medium is given by 
the proper dispersion relation determined from the solution to
\begin{equation}
\label{photondisp}
\omega^2(\kappa) -  \kappa^2 =
\pi_{\!_T}(\omega(\kappa),\kappa)\,.
\end{equation}
A common practice is to express $\omega(\kappa)$ in terms of the
index of refraction  $n\equiv\kappa/\omega(\kappa)$. The quantity of interest is 
the  energy $\mathcal{S}$ radiated forward, into the
vacuum. This is computed from the transition probability $\mathcal{W}$ for the process 
in Eq. (\ref{process})
\begin{equation}
d\mathcal{W} = \frac{V d^3\wp'}{(2 \pi)^3} \frac{V d^3
\kappa'}{(2\pi)^3} {\lvert  S_{fi} \rvert}^2,
\end{equation}
where
\begin{align}
\label{matrixelement}
&{\lvert S_{fi} \rvert}^2  = \,\frac{\pi^3}{vV^2}
\frac{{\lvert
\mathcal{M} \rvert}^2}{{\mathcal E}{\mathcal E}'\omega}
\,\delta\!\left( \wp_x - \wp'_x- \kappa_x \right) \delta\!\left(
\wp_y -
\wp'_y- \kappa_y \right)\nonumber\\
&\times \delta\!\left( {\mathcal E} - {\mathcal E}^{\prime}- \omega
\right) \left\lvert{\int_{-\ell/2}^0\mathrm{d}z \exp[i\!\left(\wp -
\wp'_z- \kappa_z \right)\!z]}\right\rvert^2 \! .
\end{align}
Here,  $V ={\ell}^{\,3}$ denotes the volume of the transition region and 
$v = \ell/\tau$ is the neutrino velocity expressed in terms of
the time interval of the process $\tau$. In writing Eq. (\ref{matrixelement}) 
we used the fact that,  for the neutrino-photon 
interaction in matter, the matrix element $S_{fi}$ vanishes outside of the
medium ($z>0$).

The emitted photon is described by a monochromatic wave with a definite 
(transverse) polarization vector 
$\epsilon_{\mu}(k,\l) = (0, \boldsymbol{\epsilon}(k,\l))$ ($\l = 1,2$), which satisfies 
$\epsilon(k,\l)\cdot k
=\epsilon(k,\l)\cdot u = 0$ and
$\epsilon(k,\l)\cdot\epsilon(k,\l')=\delta_{\l\l'}$.
Thus, for the $\nu\nu\gamma$  amplitude we have
\begin{equation}
\label{vertex} \mathcal{M} = -i
\sqrt{\!\mathcal{N}}\,\bar{u}(p^{\prime})
\Gamma^{\mu}u(p)\epsilon_{\mu}(k,\l)\,,
\end{equation}
where $u(p)$ represents the Dirac spinor with momentum $p$. The factor
$\sqrt{\!\mathcal{N}}$ has to be included because the normalization
of the photon wave function in the medium differs from the one in
vacuum \cite{zaidi}. The electromagnetic vertex function $\Gamma^{\mu}$ refers to the
background part and depends not only on the momenta of the ingoing
and outgoing neutrino, but also on $u^\mu$. 

\begin{figure}[h]
\label{figGeom}
\includegraphics[viewport=0 0 260 180, width = 8 cm]{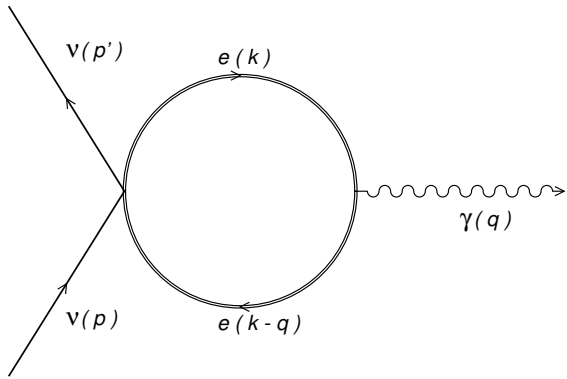}
\caption{One-loop diagram contributing, in the local approximation, to the neutrino-photon vertex in matter.
The internal double lines represent the electron thermal propagators.}
\end{figure}

In many circumstances electrons make the dominant contribution to the
neutrino  electromagnetic vertex and, in a first approximation, 
we ignore the contributions  coming from the protons and neutrons present in 
the background \footnote {The nucleon contributions to the neutrino electromagnetic vertex in matter, 
including the effect of the anomalous moment of the  nucleons, has been calculated in Ref. \cite{dn}.
They should be incorporated in a more general treatment of the problem.}.
The general expression for $\Gamma^{\mu}$  in a medium consisting of an electron gas
has been explicitly calculated to lowest order in the Fermi coupling constant $G_F$ \cite{sem}.
In the local approximation, i.e, neglecting the momentum dependence in the weak bosons propagators 
the contributions to the $\nu \nu \gamma$ vertex reduce to the single diagramm shown in Fig. 2. 
From the result presented in the works of Ref. \cite{sem} we can write
\begin{equation}
\label{emvertex}
\Gamma_{\mu}=-\sqrt{2}\, \frac{G_F}{e}
(a\,\pi_{\mu\nu}+b\,\pi^A_{\mu\nu})\gamma^\nu L\,,
\end{equation}
where $L=\frac{1}{2}(1-\gamma_5)$ and $e$ is the electric charge of
the electron. In the SM
$a = 2 \sin^2 \theta_W \pm \frac{1}{2}$ and $b = \mp\frac{1}{2}$,
where the upper sign corresponds to the $\nu_e$ and the lower sign
to the $\nu_{\mu,\tau}$.  The first and second term in Eq. \eqref {emvertex} 
correspond to the contribution coming from the vector and axial vector 
part of the electron current, respectively, in the effective four-fermion 
interaction between electrons and neutrinos.

The formula for  $\pi_{\mu \nu}$ is the one given in  Eq. \!(\ref{fotonselfe}), while
$\pi^A_{\mu\nu} =  \pi_{\!_A}(\omega,\kappa) P_{\mu\nu}\hs{.02cm},$
with $P_{\mu\nu} =
i\,\epsilon_{\mu\nu\alpha\beta} k^\alpha u^\beta/\kappa\,.$ It can be immediately 
verified that $k^\mu\pi_{\mu\nu}=k^\mu\pi^5_{\mu\nu}=0$, which in turn implies that
$k^\mu\Gamma_\mu =0$ as required by gauge invariance. At this point, 
it is pertinent to note that the  most general expression for  $\pi_{\mu\nu}$ includes 
a term proportional to the tensor $P_{\mu\nu}$. For a  parity-conserving media such 
a term can arise only through higher-order contributions to the photon self-energy 
involving the weak interaction of the particles present in the background and hence 
must be small \cite{nipal}. Consequently, we ignored it in writing Eq. (\ref{fotonselfe}), 
thus taking the transverse modes degenerated.

Performing the integrations on $d^3\wp'$ the energy radiated into the vacuum becomes
\begin{equation}
\label{enint}
 \frac{d^2 \mathcal{S}}{d\omega d\t'} = \omega
\frac{d^2 \mathcal{W}}{d\omega d\t'} = \frac{1}{32\pi^2v} \,
\frac{\omega^2\lvert \overline{\mathcal{M}} \rvert^2\sin\t'}
{{\mathcal E}\wp'_z\!\left({\kappa}_z + \wp'_z-\wp \right)^2}\, ,
\end{equation}
with $\kappa_z =  \kappa \cos\t$ and
$\wp'_z  =  \sqrt{\wp^2 - 2{\mathcal E} \omega + \omega^2 \cos^2
\t'}\,$. Here, $\t$ and $\t'$ are the angles that the photon momentum 
makes with the $z$ axis within and outside the medium, respectively. 
They are related by the Snell's law $\sin\t' = n \sin\t$ ($n' =1$), 
as a result of conservation of the transverse photon momentum 
($\kappa_{x,y} = \kappa'_{x,y}$). For $n$ close
to 1, the distinction between $\t$ and $\t'$ is not 
numerically significant, but it has to be kept to avoid a 
spurious singularity in the angular integration
From a physical point of view, this is related to the fact that
photons are emitted into the vacuum at angles smaller than $\pi/2$.

The quantity $\lvert \overline{\mathcal{M}} \rvert^2$ is calculated by 
averaging over the initial neutrino spins and summing over the neutrino 
final spins and the two photon polarizations. The transverse and axial polarization 
functions $\pi_{_T}$ and $\pi_{\!_A}$ are evaluated at the photon dispersion  relation in 
the medium \footnote {Momentum is conserved at the interaction vertex, even 
though it is not in the whole process since, when photons 
are radiated into the vacuum, part of the momentum of the initial neutrino is taken 
up by the medium.}. They are given by integrals over the momenta of the 
electrons and positrons in the background. The most comprehensive analytical 
expressions for these functions are those given by Braaten and Segel \cite{braaten}. 
From them, one can show that in the classical, degenerate, and relativistic limit
$\pi_{\!_{A}}/\pi_{_T} \lesssim  \sqrt{\alpha/3\pi}$, with $\alpha=e^2/4\pi$.  
As a result, we discard the contributions to $\lvert \overline{\mathcal{M}} \rvert^2$ 
that come from the axial vector part in $\Gamma_\mu$. This leads to
\begin{equation}
\label{Mbar} 
\lvert\overline{\mathcal{M}}\rvert^2 \cong  \mathcal{N}
\frac{G_F^2}{\pi \alpha}\,a^2 \lvert \pi_{\!_T} \rvert^2
\negmedspace \,\big({\mathcal E} {\mathcal E}'\! - \wp\,\wp'_z \cos^2\t
+\kappa \wp\cos\t\sin^2\t\big)\,,
\end{equation}
where we took into account that $Q^{\mu\nu}\epsilon_{\mu}(k,\l) =0$
and used the relation $\sum_{\l=1,2}\epsilon_{\mu}(k,\l) 
\epsilon_{\nu}(k,\l)  = -R_{\mu\nu}$ valid for the polarization
vectors in the medium.  

The energy spectrum is obtained by substituting (\ref{Mbar}) into (\ref{enint}) 
and integrating over the angular variable. The upper value for the angle in vacuum 
depends on $\omega$ and it is simpler to integrate over the angle in the medium, 
for which $0\leqslant \t \leqslant \pi/2$. Then, using $d\t' / d\t = n \cos\t/\sqrt{1-n^2 \sin\t}$, 
for a relativistic neutrino (${\mathcal E}\cong\wp)$ we find
\begin{equation}
\label{dSdo}
\frac{d\mathcal{S}}{d\omega} =
\frac{\mathcal{N}}{v}\,\frac{G_F^2\,a^2}{32\pi^3 \alpha}\,\kappa^2
\lvert \pi_{\!_T} \rvert^2 \mathcal{F}(\omega)\,,
\end{equation}
where
\begin{equation}
\label{F} 
\mathcal{F}(\omega)=\int^1_0\!\frac{\mathrm{d}\z
\,\z} {\sqrt{1-n^2(1-\z^2)}}
\left[ \frac{\wp - \omega - \wp'_z \z^2+\kappa\hs{.03cm}
\z(1-\z^2)}{\wp'_z\!\left(\wp'_z + \kappa \hs{.03cm} \z - \wp \right)^2}\right]\,,
\vspace{0.2cm}
\end{equation}
with $\z = \cos\t$. In writing Eq. (\ref{F}) we took into consideration 
Snell's law and that in vacuum $\kappa' = \omega$.  

The last integral can not be done in a closed form, however a good 
analytic result for $\mathcal{F}(\omega)$ is obtained by putting $f(\z) = \sqrt{1-n^2(1-\z^2)}$ 
equal to one in the integrand. In fact, $f(\z)$ differs significantly from unity only 
for $n\approx 1$, but in this case the factor between square brackets exhibits a 
sharp maxima at $\z \approx 1$.  Since $f(1) =1$ we can replace
$f(\z)$ by one within the integral. Proceeding in this way we arrive to
\begin{equation}
\label{dSdo2}
\frac{d\mathcal{S}}{d\omega} = 
\frac{\mathcal{N}}{v}\,\frac{G_F^2\,a^2}{64\pi^3 \alpha}\,\lvert \pi_{\!_T} \rvert^2\!
\left[\!\left(1-\frac{\omega}{\wp}\right)\!\mathcal{I}(s) 
-\,\frac{\wp^2}{2\kappa^2}\mathcal{L}(s)
+ \frac{\wp^2}{4\kappa^2}\!\left(\frac{2\kappa^2}{\wp^2}
- 1 + \varrho \right)\!\mathcal{J}\!(s)\right]^{s_1}_{s_0},
\end{equation}
where $\varrho=[(\wp-\omega)^2-\kappa^2]/\wp^2$, $s_1 = (\wp-\omega+\kappa)/\wp$,
and $s_0=\lvert \varrho \rvert^{1/2}$. The functions $\mathcal{I}(s)$, $\mathcal{J}(s)$, and $\mathcal{L}(s)$
are given by
\begin{eqnarray}
\mathcal{I}(s)&=&\frac{1-\varrho}{1-s} +
\frac{\varrho}{s}+2\varrho\ln\hs{.02cm}(1-s)- 2\varrho\ln s\,,\nonumber \\
\mathcal{J}\!(s)&=&\frac{(1-\varrho)^2}{1-s}-\dfrac{\varrho^2}{2s^2}(4s +1)
+\,(3\varrho +1)(1-\varrho)\ln\hs{.02cm}(1-s) + (3\varrho -2)\varrho \ln s\,,\nonumber\\
\mathcal{L}(s)&=&\frac{s}{4}(s+4)+\dfrac{\varrho^2}{4s^2}(4s +1)
+ \hs{0.02cm} (1-\varrho)^2\ln\hs{.02cm}(1-s) + (1-\varrho)\varrho\ln s\,.
\end{eqnarray}
It is important to remark that the formula (\ref{dSdo2}) has been derived without any 
assumption about the background. It can be applied to compute the energy emitted 
by a relativistic  neutrino when it crosses the interface between the vacuum and an electron gas of
any type. 

From the expression of $\mathcal {N}$ as a function of $\omega$  
\cite{braaten} one can show that it is always close to unit and in the numerical calculations
we take $\mathcal{N} = 1$. On the other hand, the momentum $\kappa$ has to be expressed in 
terms of the photon energy by solving Eq. (\ref{photondisp}). According to the results 
of Ref. \cite{braaten} this entails to finding the solution of the transcendental equation
\begin{equation}
\label{disprel} 
\omega^2-\kappa^2 =
\omega_p^2\left[1+\frac{1}{2}G(v_*^2\kappa^2/\omega^2)\right],
\end{equation}
where $ v_* \leq 1$ is a ``typical" velocity of the electrons
in the plasma 
and the function $G$ is defined by
\begin{equation}
G(x) = \frac{3}{x} \left[ 1 - \frac{2 x}{3} -
\frac{(1-x)}{2\sqrt{x}}\log\frac{1+\sqrt{x}}{1-\sqrt{x}}\right],
\quad 0 \leq x \leq 1.
\end{equation}
A convenient simplification can be introduced by means of an 
iterative procedure. First, we solve Eq. (\ref{disprel}) by expanding function $G$ 
in a power series and keeping the lowest order term. This gives us ${\kappa^2}/{\omega^2} =
({\omega^2 - \omega_p^2}) / ({\omega^2 +\frac{1}{5} {v_*^2 \omega_p^2}})$
that corresponds to the first relativistic correction to the photon dispersion relation. 
Next, we make  ${\kappa^2}/{\omega^2}$ equal to this approximate value in
the argument of $G$ on the right-hand side of Eq. (\ref{disprel}), which
renders us $\kappa$ as a new explicit function of $\omega$:
\begin{equation}
\kappa^2 \cong \omega^2 - \omega_p^2\left[1 +\frac{1}{2}G\left(\upsilon_*^2 \,\frac {\omega^2 - \omega_p^2}
{\omega^2 + \frac{\upsilon_*^2 \omega_p^2}{5}}\right)\right].
\end{equation}
This function works remarkably well in the whole
range of frequencies and we use it to perform the
numerical integration, over the interval  $(\omega_p, \wp)$, 
involved in the calculation of the total energy $\mathcal{S}$
radiated into the vacuum.

\begin{figure}[htb]
\includegraphics[width=14cm,viewport= 50 50 554 338]{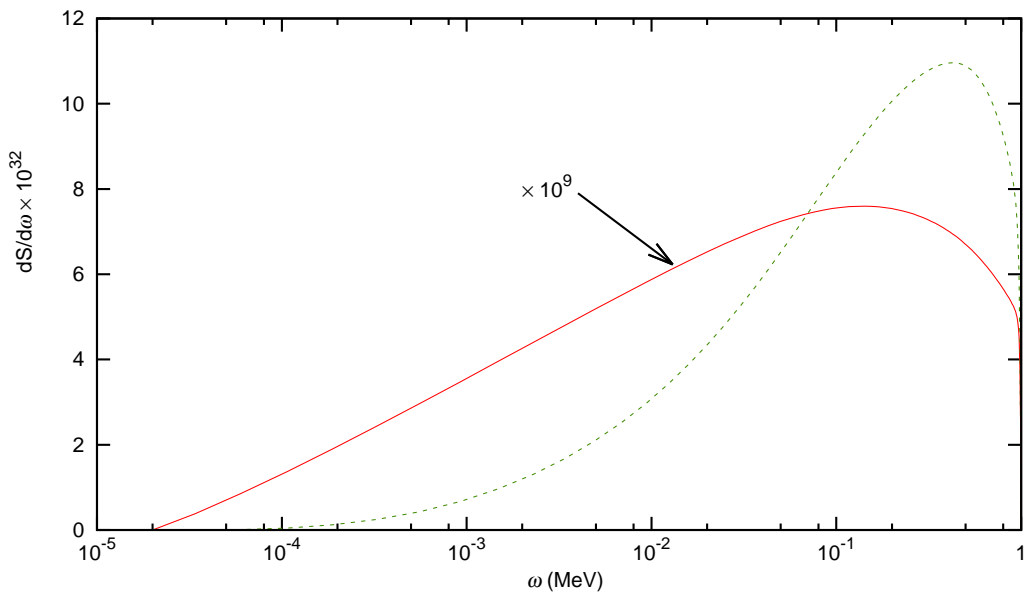}
\caption{Energy spectrum of the TR from the neutrino-photon 
coupling in matter for an incident neutrino with an energy 
${\mathcal E}= 1 \mathrm{ MeV}$. The solid red and the dashed green lines 
correspond to a classical electron gas with $\omega_p = 20\,\mathrm{eV}$ 
and  a degenerate gas with $\omega_p = 5\,\mathrm{KeV}$, respectively.}
\end{figure}

In Fig. 3 we have plotted the energy emitted by a 
$1 \, \mathrm{MeV}$ incident $\nu_e$ versus the photon 
energy, for two limiting situations: (i) a classical electron gas with $\omega_p = 20\,
\mathrm{eV}$ (polypropylene) at room temperature 
($v_* = \sqrt{5T/m_e} \cong 0$) and (ii) a degenerate gas 
($v_* = v_{_{F}} = 0.3$)
with $\omega_p = 5\,\mathrm{KeV}$ 
\footnote{ For a degenerate gas $\omega_p = (4\alpha/3\pi) p_{_F}^2v_{_F}$ and
$n_e = p_{_F}^3/3\pi$, where $v_{_F} = p_{_{ F}}/E_{_{F}}$
and $n_e$ is the electron number density.}.
As clearly  illustrated in this figure,  the intensity of the phenomenon and
the shape of the spectrum depend markedly on the properties of the medium.
For a degenerate plasma the TR energy is much higher and the spectrum is
more sharp than in the classical limit.  For the classical gas the total radiated 
energy  in a single interface is $\mathcal{S} \cong 6.6 \times10^{-35} \,\mathrm{eV}$,
which is two orders of magnitude larger than the value obtained in the 
case of  a toroidal dipole moment  \cite{bukina}. 

It is interesting to relate our result with the one due to a neutrino magnetic 
moment $\mu_{\nu}$. From Eq. (13b) 
of Ref. \cite{sakuda} 
\begin{equation}
\mu_\nu = 1.5 \times 10^{6} \mu_B \sqrt{\frac{ \mathcal{S}_{\!_M}}{\mathcal{E}}}\,,
\end{equation}
where $\mu_B$ is the Bohr magneton and $\mathcal{S}_{\!_M}$ denotes the total energy radiated 
because of $\mu_{\nu}$. When evaluated in $\mathcal{S}_{\!_M} = 6.6 \times10^{-35} \,\mathrm{eV}$
and $\mathcal{E} = 1 \mathrm{MeV}$, the above formula yields $\mu_\nu \cong 1.2 \times 10^{-14} \mu_B$, 
which is several order of magnitude larger than the prediction of the minimally
extended SM ($\mu_\nu \approx 3 \times 10^{- 19} \mu_B  [m_\nu/1\mathrm{eV}])$. 
The energy radiated by a magnetic moment  increases, and becomes 
higher than the one predicted here,  if we take a larger value for $\mu_{\nu}$, 
close to the present experimental limits. However, two things should kept in mind:
i) such limits might be quite poor, as astrophysical arguments seem to indicate \cite{raffelt,adt}, and
ii) the process under consideration is a real one, in the sense that it represents a firm 
prediction based solely on the physics of the SM.  Moreover, a $\mu_{\nu} \sim 10^{-14}\mu_B $
is of the same order as to the upper bound on the magnetic moments of Dirac neutrinos 
generated by physics above the scale of electroweak symmetry breaking \cite{bell}. 

A practical TR detector consists of several sets of a radiator, with a stack of thin 
foils of  a certain material, to produce the TR photons and a gas proportional chamber 
to detect them. In the works of  Refs. \cite{sakuda} and \cite{bukina} the authors consider 
a detector of 10 m$^2$ area made up of 10 sets of radiators and xenon chambers, 
each radiator comprising $10^4$ polypropylene foils. In our case, for the same experimental 
setup and a flux of $10^{13}$cm$^{-2}$s$^{-1}$ antineutrinos coming from a nuclear reactor, we
get $W = 2.1 \times 10^{-4}$ [$\mathcal {T}$/1year] eV, with $W$ being the total energy deposited during 
a time interval $\mathcal{ T}$. Such a small value precludes its possible experimental observation. 
The measurement feasibility could be improved by enlarging the detector, for example, 
augmenting the number of foils. Finally, let us notice (see Eqs. \eqref{dSdo2} and \eqref{disprel})
that the radiated energy  goes like $\omega_p^4$ and, as indicated before,   
increases enormously if, instead of a classical gas, we consider 
a degenerate electron plasma  similar to those existing in stellar objects. The calculations
 presented in this paper are a first step towards a full understanding of this radiative neutrino 
 phenomena and serve as a useful framework for further studies of its implications in such 
 astrophysical environments and
other physical situations.
\

\begin{acknowledgments}
The authors acknowledge support by DGAPA-UNAM under Grant No. PAPIIT IN117210 
and by CONACyT under Grant No. 83534 and Red FAE.
\end{acknowledgments}


\begin{thebibliography}{99}


\bibitem{giunti} For a recent review see C. Giunti and A. Studenikin,
 Phys. Atom. Nucl. 32, 2089 (2009). 
  
\bibitem{raffelt} G. Raffelt, Phys. Rep. 320,  319 (1999). 

\bibitem{bell} N. F. Bell et al., Phys. Rev. Lett. 95, 151802 (2005); 
N. Bell, Int. J. Mod. Phys. A 22, 4891 (2007).

\bibitem{novales} H. Novales-S\'anchez, A. Rosado, 
V. Santiago-Ol\'an, and J. J. Toscano,  Phys. Rev. D78, 073014 (2008).

\bibitem{diracmaj} J. F. Nieves, Phys. Rev. D 26, 3152 (1982); 
B. Kayser, Phys. Rev. D 26, 1662 (1982); Phys. Rev. D 30, 1023 (1984). 

\bibitem{sem} V. N. Oraevski\u{\i}, V. B. Semikoz, and Ya. A. Smorodinski\u{\i}, 
JETP Lett. 43, 709 (1986); V. N. Oraevski\u{\i}, A. Yu. Plakhov, V. B. Semikoz, 
and Ya. A. Smorodinski\u{\i}, Sov. Phys. JETP 66, 890 (1987); Erratum ibid. 68, 1309 (1989); 
J. C. D'Olivo, J. F. Nieves, and P. B. Pal, Phys. Rev. D 40, 3679 (1989).

\bibitem{dn} J. C. D'Olivo and J. F. Nieves, Phys. Rev. D 56, 5898 (1997).

\bibitem{plasmon}  J. B. Adams, M. A. Ruderman, and C. H. Woo, Phys. Rev.
129, 1383 (1963); G. Beaudet, V. Petrosianãand E. E. Salpeter, 
Astrophys. J. 150, 979 (1967).

\bibitem{zaidi} M. H. Zaidi, Nuovo Cimento A 40, 502 (1965).

\bibitem{tsytov} V. N. Tsytovich, Sov. Phys. JETP 18, 816 (1964).

\bibitem{radiative} J. C. D'Olivo, J. F. Nieves, and P. B. Pal, 
Phys. Rev. Lett. 64, 1088 (1990); D. Grasso and V. Semikoz, 
Phys. Rev. D 60, 053010 (1999).

\bibitem{sawyer} R. F. Sawyer, Phys. Rev. D 46, 1180 (1992). 

\bibitem{cherneu} J. C. D'Olivo, J. F. Nieves, and P. B. Pal,
Phys. Lett. B 365, 178 (1996). 

\bibitem{frankg} V. L. Ginzburg and I. Frank, J. Phys. (USSR)
9, 353 (1945); V. L. Ginzburg and V. N. Tsytovich, {\it Transition
Radiation and Transition Scattering}, (Adam Hilger, New York,
1988). 

\bibitem{sakuda} M. Sakuda, Phys. Rev. Lett. 72 804 (1994);
M. Sakuda and Y. Kurihara, Phys. Rev. Lett. 74,1284 (1995).

\bibitem{grimus}
W. Grimmus and H. Neufeld, Phys. Lett. B 344,  252 (1995).

\bibitem{bukina} E. N. Bukina, V. M. Dubokiv, and V. E. Kuznetsov, Phys.
Lett B 435, 134 (1998).

\bibitem{doloza} J. C. D'Olivo and J. A. Loza, {\it Transition Radiation of Neutrinos}. 
Talk presented  (JAL) at the VIII Latin American Symposium on 
High Energy Physics  (SILAFAE 2010), Valpara\'{\i}so, Chile, December 6-12, 2010. 
(Unpublished).

\bibitem{weldon} H. A. Weldon, Phys. Rev. D 26, 1394 (1982).

\bibitem{nipal} J. F. Nieves and P. B. Pal, Phys. Rev. D 39, 652 (1989);
Erratum-ibid. D 40, 2148 (1989).

\bibitem{braaten} E. Braaten and D. Segel, Phys. Rev. D 48, 1478
(1993).

\bibitem{adt} A. Ayala, J C. D'Olivo, and M. Torres, Phys. Rev. D59, 11190 (1999).

\end{thebibliography}
\end{document}